\pdfoutput=1
\documentclass[a4paper]{PoS}
\usepackage{amsmath}
\usepackage{axodraw2}
\usepackage{microtype}
\usepackage{physics}

\allowdisplaybreaks

\usepackage{xfrac}
\usepackage{xspace}

\newcommand{\eps}{\epsilon}
\newcommand{\D}{\textrm{d}}
\newcommand{\PS}{\textrm{PS}}
\newcommand{\nn}{\nonumber}
\newcommand{\x}{\,}

\newcommand{\qgraf}{\texttt{QGRAF}\xspace}
\newcommand{\form}{\texttt{FORM}\xspace}
\newcommand{\fire}{\texttt{FIRE5}\xspace}
\newcommand{\litered}{\texttt{LiteRed}\xspace}
\newcommand{\dream}{\texttt{DREAM}\xspace}
\newcommand{\summertime}{\texttt{SummerTime}\xspace}

\def\z(#1){
 \zeta_{#1}
}

\def\num(#1){
  \begin{axopicture}(95,55)(2,-2)
  \textcolor[rgb]{0.4,0.4,0.4}{\PText(86,0)(0)[c]{\footnotesize $$}}
}
\def\dot(#1,#2){
  \White{\Vertex(#1,#2){3}}
}
\def\den(#1,#2,#3,#4){
  \Line[color=BrickRed,width=0.85](#1,#2)(#3,#4)
  \Vertex(#1,#2){1.5}
  \Vertex(#3,#4){1.5}
}
\def\cut(#1,#2,#3,#4){
  \Line[color=Gray,width=0.5,dash](#1,#2)(#3,#4)
  \Vertex(#1,#2){1.5}
  \Vertex(#3,#4){1.5}
}
\def\cutd(#1,#2,#3,#4){
  \DoubleLine[color=Gray,width=0.5,dash](#1,#2)(#3,#4){1.5}
  \Vertex(#1,#2){1.5}
  \Vertex(#3,#4){1.5}
}
\def\arc(#1,#2,#3,#4,#5){
  \Arc[color=Gray,width=0.5,dash](#1,#2)(#3,#4,#5)
}

\newcommand{\url}[1]{\href{#1}{#1}}

\title{
    \hfill{\small\texttt{DESY 18-136}} \\~\\
    Five-Particle Phase-Space Integrals in QCD\footnote{Based on~\cite{GMP18}.}
}

\ShortTitle{Five-Particle Phase-Space Integrals in QCD}

\author{\speaker{Oleksandr~Gituliar}\\
        II. Institut f\"ur Theoretische Physik, Universit\"at Hamburg\\
        E-mail: \email{oleksandr@gituliar.net}}

\author{Vitaly~Magerya\\
        II. Institut f\"ur Theoretische Physik, Universit\"at Hamburg\\
        E-mail: \email{vitaly.magerya@tx97.net}}

\author{Andrey~Pikelner\\
        II. Institut f\"ur Theoretische Physik, Universit\"at Hamburg\footnote{
            On leave of absence from Joint Institute for Nuclear Research, Dubna, Russia.}\\
        E-mail: \email{andrey.pikelner@desy.de}}

\abstract{
We present analytical expressions for the 31 five-particle phase-space master
integrals in massless QCD as an $\eps$-series with coefficients being multiple
zeta values of weight up to 12. In addition, we provide a computer code for the
Monte-Carlo integration in higher dimensions, based on the RAMBO algorithm,
that has been used to numerically cross-check the obtained results in 4, 6, and
8 dimensions.}

\FullConference{Loops and Legs in Quantum Field Theory (LL2018)\\
		29 April 2018 - 04 May 2018\\
		St. Goar, Germany}

\begin{document}

\section{Introduction}

Nowadays, perturbative calculations play the key role in describing data from
high-energy particle colliders, such as the LHC, as well as in improving the
precision of numerical parameters in the Standard Model and other models. It is
clear now that higher-order calculations will play an even more crucial role in
processing data from future high-luminosity colliders, like the FCC or the ILC,
where theoretical errors will dominate over experimental statistical errors.
These arguments motivate us to make one step forward beyond available fully-inclusive
phase-space integrals for a four-particle decay~\cite{GGH03} and
calculate a set of yet unknown integrals that corresponds to a five-particle
decay of a color-neutral off-shell particle in Quantum Chromodynamics.

A particular application of these integrals we have in mind is
the extraction of NNLO time-like splitting functions~\cite{AMV11}
from a semi-inclusive one-particle decay process, as for example
discussed in~\cite{Git15, GM15}: the integrals we will be looking
at correspond to a fully inclusive cross section, and can be used
to determine integration constants when calculating exclusive
quantities using the method of differential equations.

In this article, we focus on the calculation of master
integrals that can be used to express any other integral of the
corresponding topology provided a set of integration-by-parts rules
(IBP)~\cite{CT81} is known. Our approach is based on techniques
for solving dimensional recurrence relations (DRR)~\cite{Tar96}
described in~\cite{LM17a,LM17b}. In particular, we use the \dream
package~\cite{LM17a} to obtain numerical results for the desired
integrals with 2000-digit precision, and restore their analytical
form in terms of multiple zeta values (MZV)~\cite{Fur03,BBV09,LM15}
up to weight 12 using the PSLQ method~\cite{FBA99} as implemented
in Mathematica. We also present a Monte-Carlo code, based on the
RAMBO algorithm~\cite{KSE85}, for numerical integration of the
phase-space integrals in arbitrary (integer) number of dimensions
that has been used to check consistency of the obtained results.

This article is organized as following. In Section~\ref{sec:2}~we introduce
our notation and describe our calculational method in more detail. In
Section~\ref{sec:4}~we provide complete results for four-particle integrals
and discuss numerical cross-checks using Monte-Carlo integration. In
Section~\ref{sec:5}~we make our final remarks.

Accompanying to this paper are the auxiliary files on
the arXiv\footnote{\url{https://arXiv.org/abs/1803.09084}} containing
the complete master integrals with MZV weight up to 12, as well
as the Monte-Carlo integration routines with the corresponding
results.

\section{The Method}
\label{sec:2}

\begin{table}
\footnotesize
\begin{tabular}{c c c c}
\num(1)
\Line(10,20)(20,20)
\Line(80,20)(90,20)
\arc(50,10,31,20,160)
\arc(50,-15,46,50,130)
\arc(50,55,46,-130,-50)
\arc(50,30,31,-160,-20)
\cut(20,20,80,20)
\end{axopicture}
 & \num(2)
\Line(10,20)(20,20)
\Line(80,20)(90,20)
\den(20,20,40,40)
\cut(40,40,60,40)
\den(60,40,80,20)
\den(20,20,40,0)
\cut(40,0,60,0)
\den(60,0,80,20)
\cut(40,40,60,0)
\cut(60,40,40,0)
\cut(20,20,80,20)
\end{axopicture}
 & \num(3)
\Line(10,20)(20,20)
\Line(80,20)(90,20)
\den(20,20,50,40)
\cut(50,40,80,20)
\cut(20,20,50,0)
\den(50,0,80,20)

\arc(30,20,27,-45,45)
\arc(70,20,27,135,215)
\cut(50,40,50,0)
\end{axopicture}
 & \num(4)
\Line(10,20)(20,20)
\Line(80,20)(90,20)
\den(20,20,40,40)
\den(40,40,60,40)
\cut(60,40,80,20)
\cut(20,20,40,0)
\den(40,0,60,0)
\den(60,0,80,20)
\cutd(40,40,60,0)
\cut(60,40,40,0)
\end{axopicture}

\\ $F_{1}$ & $F_{2}$: 12 14 23 34 & $F_{3}$: 01 05 & $F_{4}$: 12 13 02 03
\\
\num(5)
\Line(10,20)(20,20)
\Line(80,20)(90,20)
\den(20,20,40,40)
\den(40,40,60,40)
\cut(60,40,80,20)
\cut(20,20,40,0)
\den(40,0,60,0)
\den(60,0,80,20)
\cut(40,40,40,0)
\cut(60,40,60,0)
\cut(40,40,60,0)
\end{axopicture}
 & \num(6)
\Line(10,20)(20,20)
\Line(80,20)(90,20)
\den(20,20,40,40)
\den(40,40,60,40)
\cut(60,40,80,20)
\cut(20,20,40,0)
\den(40,0,60,0)
\den(60,0,80,20)
\den(40,40,40,20)
\cut(40,20,40,0)
\cut(60,40,60,20)
\den(60,20,60,0)
\cut(40,20,60,20)
\end{axopicture}
 & \num(7)
\Line(10,20)(20,20)
\Line(80,20)(90,20)
\den(20,20,50,40)
\cut(50,40,80,20)
\cut(20,20,50,0)
\den(50,0,80,20)

\cut(20,20,80,20)
\dot(50,20)
\cutd(50,40,50,0)
\end{axopicture}
 & \num(8)
\Line(10,20)(20,20)
\Line(80,20)(90,20)
\den(20,20,40,40)
\cut(40,40,60,40)
\den(60,40,80,20)
\cut(20,20,40,0)
\cut(40,0,60,0)
\cut(60,0,80,20)
\den(40,40,60,0)
\cut(20,20,80,20)
\dot(50,20)
\den(60,40,40,0)
\end{axopicture}

\\ $F_{5}$: 12 34 02 03 & $F_{6}$: 12 13 02 04 45 35 & $F_{7}$: 123 124 & $F_{8}$: 13 14 123 124
\\
\num(9)
\Line(10,20)(20,20)
\Line(80,20)(90,20)
\den(20,20,40,40)
\cut(40,40,60,40)
\den(60,40,80,20)
\den(20,20,40,0)
\cut(40,0,60,0)
\den(60,0,80,20)
\cut(40,40,60,0)
\cutd(60,40,40,0)
\end{axopicture}
 & \num(10)
\Line(10,20)(20,20)
\Line(80,20)(90,20)
\cutd(20,20,50,40)
\den(50,40,80,20)
\den(20,20,50,0)
\cut(50,0,80,20)

\cutd(50,40,50,0)
\end{axopicture}
 & \num(11)
\Line(10,20)(20,20)
\Line(80,20)(90,20)
\den(20,20,40,40)
\den(40,40,60,40)
\cut(60,40,80,20)
\cut(20,20,40,0)
\den(40,0,60,0)
\den(60,0,80,20)
\cut(40,40,60,0)
\cutd(60,40,40,0)
\end{axopicture}
 & \num(12)
\Line(10,20)(20,20)
\Line(80,20)(90,20)
\den(20,20,50,40)
\cut(50,40,80,20)
\cut(20,20,40,0)
\den(40,0,60,0)
\den(60,0,80,20)
\cut(50,40,40,0)
\cut(50,40,60,0)
\cut(60,0,20,20)
\end{axopicture}

\\ $F_{9}$: 12 23 023 012 & $F_{10}$: 01 123 & $F_{11}$: 01 123 014 05 & $F_{12}$: 12 023 05
\\
\num(13)
\Line(10,20)(20,20)
\Line(80,20)(90,20)
\den(20,20,30,35)
\den(30,35,50,40)
\den(50,40,70,35)
\cut(70,35,80,20)
\den(70,5,80,20)
\den(50,0,70,5)
\cut(50,0,70,35)
\cut(30,35,50,0)
\cut(70,5,50,40)
\cut(20,20,70,5)
\end{axopicture}
 & \num(14)
\Line(10,20)(20,20)
\Line(80,20)(90,20)
\cut(20,20,40,40)
\cut(40,40,60,40)
\cut(60,40,80,20)
\den(20,20,40,0)
\cut(40,0,60,0)
\den(60,0,80,20)
\den(40,40,60,0)
\cut(40,0,40,40)
\dot(50,20)
\den(60,40,40,0)
\end{axopicture}
 & \num(15)
\Line(10,20)(20,20)
\Line(80,20)(90,20)
\den(20,20,30,35)
\den(30,35,50,40)
\den(50,40,70,35)
\cut(70,35,80,20)
\cut(20,20,30,5)
\den(30,5,50,0)
\den(50,0,70,5)
\den(70,5,80,20)
\cut(30,35,70,5)
\cut(30,5,70,35)
\cut(50,0,50,40)
\end{axopicture}
 & \num(16)
\Line(10,20)(20,20)
\Line(80,20)(90,20)
\den(20,20,30,35)
\den(30,35,50,40)
\den(50,40,70,35)
\cut(70,35,80,20)
\cut(20,20,30,5)
\den(30,5,50,0)
\den(50,0,70,5)
\den(70,5,80,20)
\cut(30,35,70,5)
\cut(30,5,50,40)
\cut(50,0,70,35)
\end{axopicture}

\\ $F_{13}$: 12 13 124 05 02 & $F_{14}$: 02 12 023 05 & $F_{15}$: 12 13 124 134 02 03 & $F_{16}$: 12 34 02 03 124 134
\\
\num(17)
\Line(10,20)(20,20)
\Line(80,20)(90,20)
\den(20,20,50,40)
\cut(50,40,80,20)
\den(20,20,50,0)
\cut(50,0,80,20)
\cut(50,40,50,20)
\cut(50,20,50,0)
\cut(20,20,50,20)
\den(50,20,80,20)
\end{axopicture}
 & \num(18)
\Line(10,20)(20,20)
\Line(80,20)(90,20)
\cut(20,20,30,35)
\den(30,35,50,40)
\cut(50,40,70,35)
\cut(70,35,80,20)
\den(20,20,30,5)
\den(30,5,50,0)
\cut(50,0,70,5)
\den(70,5,80,20)
\cut(50,0,50,40)
\den(30,35,70,5)
\dot(50,20)
\den(30,5,70,35)
\end{axopicture}
 & \num(19)
\Line(10,20)(20,20)
\Line(80,20)(90,20)
\cut(20,20,30,35)
\cut(30,35,50,40)
\cut(50,40,70,35)
\den(70,35,80,20)
\den(20,20,30,5)
\den(30,5,50,0)
\cut(50,0,70,5)
\den(70,5,80,20)
\den(30,35,70,5)
\cut(30,5,70,35)
\dot(50,20)
\den(50,0,50,40)
\end{axopicture}
 & \num(20)
\Line(10,20)(20,20)
\Line(80,20)(90,20)
\den(20,20,30,35)
\cut(30,35,50,40)
\den(50,40,70,35)
\den(70,35,80,20)
\den(20,20,30,5)
\den(30,5,50,0)
\cut(50,0,70,5)
\den(70,5,80,20)
\cut(30,35,70,5)
\cut(30,5,70,35)
\cut(50,0,50,40)
\end{axopicture}

\\ $F_{17}$: 12 34 013 & $F_{18}$: 12 13 25 124 05 04 & $F_{19}$: 12 13 25 124 013 05 & $F_{20}$: 12 13 124 134 35 25
\\
\num(21)
\Line(10,20)(20,20)
\Line(80,20)(90,20)
\den(20,20,30,35)
\cut(30,35,50,40)
\cut(50,40,70,35)
\cut(70,35,80,20)
\den(70,5,80,20)
\den(50,0,70,5)
\cut(50,0,70,35)
\cut(30,35,50,0)
\dot(60,20)
\den(70,5,50,40)
\dot(56,31)
\dot(35,25)
\den(20,20,70,35)
\end{axopicture}
 & \num(22)
\Line(10,20)(20,20)
\Line(80,20)(90,20)
\den(20,20,30,35)
\cut(30,35,50,40)
\den(50,40,70,35)
\den(70,35,80,20)
\den(20,20,30,5)
\cut(30,5,50,0)
\cut(50,0,70,5)
\cut(70,5,80,20)
\cut(30,5,50,40)
\den(50,0,70,35)
\dot(58,14)
\dot(42,26)
\den(30,35,70,5)
\end{axopicture}
 & \num(23)
\Line(10,20)(20,20)
\Line(80,20)(90,20)
\den(20,20,30,35)
\cut(30,35,50,40)
\den(50,40,70,35)
\den(70,35,80,20)
\den(20,20,30,5)
\den(30,5,50,0)
\cut(50,0,70,5)
\den(70,5,80,20)
\cut(50,0,70,35)
\cut(30,5,50,40)
\cut(30,35,70,5)
\end{axopicture}
 & \num(24)
\Line(10,20)(20,20)
\Line(80,20)(90,20)
\den(20,20,30,35)
\den(30,35,50,40)
\cut(50,40,70,35)
\den(70,35,80,20)
\cut(20,20,30,5)
\den(30,5,50,0)
\cut(50,0,70,5)
\cut(70,5,80,20)
\cut(30,5,50,40)
\den(50,0,70,35)
\dot(42,26)
\dot(58,14)
\den(30,35,70,5)
\end{axopicture}

\\ $F_{21}$: 12 13 24 012 05 & $F_{22}$: 12 13 24 012 45 05 & $F_{23}$: 12 13 25 45 134 012 & $F_{24}$: 12 13 45 134 05 03
\\
\num(25)
\Line(10,20)(20,20)
\Line(80,20)(90,20)
\den(20,20,40,40)
\cut(40,40,60,40)
\den(60,40,80,20)
\cut(20,20,40,0)
\den(40,0,60,0)
\den(60,0,80,20)
\den(40,40,40,20)
\cut(40,20,40,0)
\cut(60,40,60,20)
\cut(60,20,60,0)
\den(40,20,60,20)
\end{axopicture}
 & \num(26)
\Line(10,20)(20,20)
\Line(80,20)(90,20)
\cutd(20,20,50,40)
\den(50,40,80,20)
\den(20,20,50,0)
\cutd(50,0,80,20)

\cut(50,40,50,0)
\end{axopicture}
 & \num(27)
\Line(10,20)(20,20)
\Line(80,20)(90,20)
\den(20,20,40,40)
\den(40,40,60,40)
\cut(60,40,80,20)
\cutd(20,20,40,0)
\den(40,0,60,0)
\den(60,0,80,20)
\cut(40,40,60,0)
\cut(60,40,40,0)
\end{axopicture}
 & \num(28)
\Line(10,20)(20,20)
\Line(80,20)(90,20)
\den(20,20,40,40)
\den(40,40,60,40)
\cut(60,40,80,20)
\cut(20,20,40,0)
\den(40,0,60,0)
\den(60,0,80,20)
\cut(40,40,40,0)
\cut(60,40,60,0)
\cut(60,40,40,0)
\end{axopicture}

\\ $F_{25}$: 12 13 45 134 012 05 & $F_{26}$: 012 123 & $F_{27}$: 12 01 123 013 & $F_{28}$: 123 012 05 01
\\
\num(29)
\Line(10,20)(20,20)
\Line(80,20)(90,20)
\cut(20,20,30,35)
\den(30,35,50,40)
\den(50,40,70,35)
\den(70,35,80,20)
\den(20,20,30,5)
\den(30,5,50,0)
\den(50,0,70,5)
\cut(70,5,80,20)
\cut(30,5,50,40)
\cut(50,0,70,35)
\cut(30,35,70,5)
\end{axopicture}
 & \num(30)
\Line(10,20)(20,20)
\Line(80,20)(90,20)
\den(20,20,30,35)
\cut(30,35,50,40)
\den(50,40,70,35)
\den(70,35,80,20)
\cut(20,20,30,5)
\cut(30,5,50,0)
\den(50,0,70,5)
\cut(70,5,80,20)
\den(30,5,50,40)
\cut(50,0,70,35)
\dot(42,26)
\dot(57,14)
\den(30,35,70,5)
\end{axopicture}
 & \num(31)
\Line(10,20)(20,20)
\Line(80,20)(90,20)
\den(20,20,40,40)
\den(40,40,60,40)
\cut(60,40,80,20)
\cut(20,20,40,0)
\den(40,0,60,0)
\den(60,0,80,20)
\cut(40,40,40,20)
\den(40,20,40,0)
\den(60,40,60,20)
\cut(60,20,60,0)
\cut(40,20,60,20)
\end{axopicture}
 & \num(32)
\end{axopicture}

\\ $F_{29}$: 12 123 25 013 05 01 & $F_{30}$: 12 13 034 134 05 03 & $F_{31}$: 12 13 034 134 05 04 &
\end{tabular}
\caption{Cut diagrams for five-particle phase-space master integrals in QCD.
Dashed lines represent cut propagators and carry final-state momenta
$p_1, \dots, p_5$. Labels represent propagators, so
that "123" corresponds to $p_1+p_2+p_3$ and
"012" to $q-p_1-p_2$ (where $q$ is the initial-state momentum, i.e., $q=p_1+\dots+p_5$).}
\label{tab:mis}
\end{table}

We start by identifying a set of five-particle phase-space master integrals
using two different approaches for consistency. As the first approach we
exploit the equivalence of IBP rules for cut and ordinary propagators,
and obtain the complete basis of phase-space master integrals by taking
all five-particle cuts of the 28 master integrals for four-loop massless
propagators found in~\cite{BC10}, discarding those that do not correspond
to the squared matrix elements of the $1 \to 5$ process (i.e.
only leaving graphs which are bipartite), and reducing the
remaining integrals with Laporta-style IBP reduction~\cite{Lap00,Lap17} as
implemented in \fire~\cite{Smi14}. As an alternative approach, we construct
the complete expression for the total cross section of the $1\to 5$ process
in QCD using \qgraf~\cite{Nog93} and \form~\cite{RUV17}, and then reduce
it with the help of \fire. Both methods give 31 master integrals listed in
Table~\ref{tab:mis}, with each having up to 6 unique propagators.
Our notation for these integrals is
\begin{equation}
    \label{eq:fi}
    F_i = S_\Gamma \int \D \PS_5 ~ \frac{1}{D^{(i)}_1 ~ \dots ~ D^{(i)}_n},
\end{equation}
where $D^{(i)}_j$ are propagators that take the form of invariant scalar
products
\begin{equation}
    s_{kl\dots q} = \left(p_k+p_l+\dots+p_q\right)^2,
\end{equation}
$\D \PS_5$ is a five-particle phase-space element in $D$ dimensions
\begin{equation}
  \label{eq:psint-def}
  \D \PS_N =
  \left (
    \prod_{i=1}^{N} \D^D p_i \x \delta^{+}\big(p_i^2\big)
  \right )
  \delta^{(D)}\Big(q - p_1 - \ldots - p_N \Big),
\end{equation}
and $S_\Gamma$ is a common normalization factor chosen for
convenience\footnote{This way we prevent additional constants (e.g, $\gamma_E$
or $\ln{\pi}$) to appear in the final results, hence reducing its size as well
as a size of the basis for PSLQ algorithm.} to be
\begin{equation}
    \label{eq:sgamma}
    S_\Gamma =
        \left(q^2\right)^{5-2D}
        \frac{\big(2\pi\big)^4}{\pi^{2D}}
        \Gamma\left(\frac{D}{2}-1\right)\Gamma\left(3\frac{D}{2}-3\right).
\end{equation}
With this normalization and knowing the volume of the complete $N$-particle
phase space\footnote{The dependence on $q^2$ is trivial here, and can be
restored by power counting. We will omit it from now on, setting $q^2$ to $1$.}
\begin{equation}
    \int \D \PS_N =
        \left(q^2\right)^{\frac{D}{2}(N-1)-N}
        \x
        \frac{\pi^{\frac{D}{2}(N-1)}}{\big(2\pi\big)^{N-1}}
        \frac{\Gamma\left(\frac{D}{2}-1\right)^N}
        {\Gamma\Big(\left(\frac{D}{2}-1\right)\big(N-1\big)\Big)
        \x
        \Gamma\Big(\left(\frac{D}{2}-1\right)N\Big)},
\end{equation}
we can already fix the value of $F_1$ as:
\begin{equation}
  \label{eq:f1}
    F_1 = S_\Gamma \int \D \PS_5 =
    \frac{\Gamma\big(\frac{D}{2}-1\big)^6 \x \Gamma\big(3\frac{D}{2}-3\big)}
    {\Gamma\big(4\frac{D}{2}-4\big) \x \Gamma\big(5\frac{D}{2}-5\big)}
\end{equation}

Next, with the help of \litered~\cite{Lee13} and \fire~\cite{Smi14} we derive
a set of lowering dimensional recurrence relations which express master
integrals in $D+2$ dimensions in terms of master integrals in $D$ dimensions:
\begin{equation}
    \label{eq:drr}
    F_i(D+2) = M_{ij}(D) \x F_j(D).
\end{equation}

In the general case $M(D)$ is expected to have a block-triangular
structure, but in our case it can be shuffled into triangular
form. The general structure of our $M(D)$ can be visualized as:

\begin{equation}
M(D)\sim
\left(
\tiny
\begin{array}{ccccccccccccccccccccccccccccccc}
     \# & - & - & - & - & - & - & - & - & - & - & - & - & - & - & - & - & - & - & - & - & - & - & - & - & - & - & - & - & - & - \\
     \# & \# & - & - & - & - & - & - & - & - & - & - & - & - & - & - & - & - & - & - & - & - & - & - & - & - & - & - & - & - & - \\
     \# & - & \# & - & - & - & - & - & - & - & - & - & - & - & - & - & - & - & - & - & - & - & - & - & - & - & - & - & - & - & - \\
     \# & - & \# & \# & - & - & - & - & - & - & - & - & - & - & - & - & - & - & - & - & - & - & - & - & - & - & - & - & - & - & - \\
     \# & - & \# & - & \# & - & - & - & - & - & - & - & - & - & - & - & - & - & - & - & - & - & - & - & - & - & - & - & - & - & - \\
     \# & - & \# & - & \# & \# & - & - & - & - & - & - & - & - & - & - & - & - & - & - & - & - & - & - & - & - & - & - & - & - & - \\
     \# & - & - & - & - & - & \# & - & - & - & - & - & - & - & - & - & - & - & - & - & - & - & - & - & - & - & - & - & - & - & - \\
     \# & - & - & - & - & - & \# & \# & - & - & - & - & - & - & - & - & - & - & - & - & - & - & - & - & - & - & - & - & - & - & - \\
     \# & - & - & - & - & - & \# & - & \# & - & - & - & - & - & - & - & - & - & - & - & - & - & - & - & - & - & - & - & - & - & - \\
     \# & - & - & - & - & - & - & - & - & \# & - & - & - & - & - & - & - & - & - & - & - & - & - & - & - & - & - & - & - & - & - \\
     \# & - & \# & - & - & - & \# & - & - & \# & \# & - & - & - & - & - & - & - & - & - & - & - & - & - & - & - & - & - & - & - & - \\
     \# & - & - & - & - & - & - & - & - & \# & - & \# & - & - & - & - & - & - & - & - & - & - & - & - & - & - & - & - & - & - & - \\
     \# & - & \# & - & - & - & - & - & - & \# & - & \# & \# & - & - & - & - & - & - & - & - & - & - & - & - & - & - & - & - & - & - \\
     \# & - & \# & - & - & - & - & - & - & \# & - & \# & - & \# & - & - & - & - & - & - & - & - & - & - & - & - & - & - & - & - & - \\
     \# & - & \# & \# & - & - & \# & \# & - & \# & \# & \# & - & \# & \# & - & - & - & - & - & - & - & - & - & - & - & - & - & - & - & - \\
     \# & - & \# & - & \# & - & \# & - & - & \# & \# & \# & - & \# & - & \# & - & - & - & - & - & - & - & - & - & - & - & - & - & - & - \\
     \# & - & - & - & - & - & - & - & - & - & - & - & - & - & - & - & \# & - & - & - & - & - & - & - & - & - & - & - & - & - & - \\
     \# & - & \# & - & - & - & - & - & - & \# & - & \# & - & \# & - & - & \# & \# & - & - & - & - & - & - & - & - & - & - & - & - & - \\
     \# & - & - & - & - & - & \# & \# & - & \# & - & \# & - & - & - & - & \# & - & \# & - & - & - & - & - & - & - & - & - & - & - & - \\
     \# & \# & - & - & - & - & \# & \# & \# & - & - & - & - & - & - & - & \# & - & - & \# & - & - & - & - & - & - & - & - & - & - & - \\
     \# & - & - & - & - & - & - & - & - & \# & - & \# & - & - & - & - & \# & - & - & - & \# & - & - & - & - & - & - & - & - & - & - \\
     \# & - & - & - & - & - & - & - & - & \# & - & \# & - & - & - & - & \# & - & - & - & \# & \# & - & - & - & - & - & - & - & - & - \\
     \# & - & - & - & - & - & \# & - & \# & - & - & - & - & - & - & - & \# & - & - & - & - & - & \# & - & - & - & - & - & - & - & - \\
     \# & - & \# & - & \# & - & - & - & - & \# & - & \# & \# & \# & - & - & \# & - & - & - & - & - & - & \# & - & - & - & - & - & - & - \\
     \# & - & - & - & - & - & \# & - & - & \# & - & \# & - & - & - & - & \# & - & - & - & - & - & - & - & \# & - & - & - & - & - & - \\
     \# & - & - & - & - & - & - & - & - & - & - & - & - & - & - & - & - & - & - & - & - & - & - & - & - & \# & - & - & - & - & - \\
     \# & - & - & - & - & - & - & - & - & \# & - & - & - & - & - & - & - & - & - & - & - & - & - & - & - & \# & \# & - & - & - & - \\
     \# & - & \# & - & - & - & - & - & - & \# & - & - & - & - & - & - & - & - & - & - & - & - & - & - & - & \# & - & \# & - & - & - \\
     \# & - & \# & \# & - & - & - & - & - & \# & - & \# & - & \# & - & - & - & - & - & - & - & - & - & - & - & \# & \# & \# & \# & - & - \\
     \# & - & \# & - & - & - & - & - & - & \# & - & \# & \# & \# & - & - & - & - & - & - & - & - & - & - & - & \# & \# & \# & - & \# & - \\
     \# & - & \# & - & - & - & - & - & - & \# & - & \# & - & - & - & - & - & - & - & - & - & - & - & - & - & \# & - & \# & - & - & \# \\
\end{array}
\right)
\end{equation}

In other words, each $F_i$ only depends on itself and master integrals from
lower sectors:
\begin{equation}
    \label{eq:drr_i}
    F_i(D+2) = \overbrace{M_{ii}(D) \x F_i(D)}^{\text{Homogeneous part}} \x + \overbrace{\sum_{k < i} M_{ik}(D) F_k(D)}^{\text{Inhomogeneous part}}
\end{equation}


As an example, for $F_{28}$ we have:
\begin{equation}
    \label{eq:drr_28}
    F_{28}(D+2) =
        \underbrace{
            -\frac{3^3}{4^3} {
                \qty(\frac{D}{2}-2)^2 \qty(\frac{D}{2}-\frac{2}{3}) \qty(\frac{D}{2}-\frac{1}{3})
                \over
                \qty(\frac{D}{2}-\frac{3}{2})^2 \qty(\frac{D}{2}-1) \qty(\frac{D}{2}-\frac{1}{2})
            }
        }_{M_{28,28}(D)}
        F_{28}(D) + \sum_{k < 28} M_{28,k}(D) F_k(D)
\end{equation}

The general solution for system~\eqref{eq:drr_i} can be written as:
\begin{equation}
  \label{eq:sol-hom-part}
  F_i(D) = \omega_i(D) \x H_i(D) + R_i(D),
\end{equation}
where
\begin{itemize}
    \item $H_i(D)$ is a homogeneous solution, which can be found
        just from the diagonal entries, $M_{ii}(D)$;
    \item $R_i(D)$ is a partial solution, which only depends on the
        integrals from subsectors, and can be constructed
        analytically as an indefinite nested sum, or evaluated
        numerically with \dream, as long as $F_1(D)$, the lowest
        integral, is known;
    \item $ \omega_i(D)$ is an arbitrary periodic function, such that
        $\omega_i(D+2)=\omega_i(D)$, which cannot be determined from
        the DRR relations alone, and needs to be fixed separately.
\end{itemize}

In the general case this solution can be evaluated as a series in $\eps$
around any dimensionality using the DRA method~\cite{Lee09},
which requires the analysis of singularities and the asymptotic
behaviour of $F_i(D)$ in the limit of $D$ going to imaginary
infinity. Our case contains two important simplifications, and
will not require such analysis.

The first simplification is that since $M(D)$ is triangular,
the homogeneous part of eq.~\eqref{eq:drr_i} decouples into a
set of first order difference equations:
\begin{equation}
    \label{eq:hi-eq}
    H_i(D+2) = M_{ii}(D)H_i(D)
\end{equation}

For rational $M_{ii}(D)$ written in the following form (compare to eq.~\eqref{eq:drr_28}):
\begin{equation}
    \label{eq:mii-rat}
    M_{ii}(D)= C\x\frac{\left(\frac{D}{2}-a_1\right)\left(\frac{D}{2}-a_2\right)\ldots\left(\frac{D}{2}-a_A\right)}{\left(\frac{D}{2}-b_1\right)\left(\frac{D}{2}-b_2\right)\ldots\left(\frac{D}{2}-b_B\right)},
\end{equation}
the homogeneous solution can immediately be found as:
\begin{equation}
    \label{eq:hi-rat-sol}
    H_i(D) = C^{\frac{D}{2}}\x{
        \Gamma\left(\frac{D}{2}-a_1\right)\Gamma\left(\frac{D}{2}-a_2\right)\ldots\Gamma\left(\frac{D}{2}-a_A\right)
        \over
        \Gamma\left(\frac{D}{2}-b_1\right)\Gamma\left(\frac{D}{2}-b_2\right)\ldots\Gamma\left(\frac{D}{2}-b_B\right)
    }
\end{equation}

Explicitly, for $H_{28}(D)$ we have:
\begin{equation}
    \label{eq:h28-rat-sol}
    H_{28}(D) = \qty(-\frac{3^3}{4^3})^\frac{D}{2}\x{
            \qty(\frac{D}{2}-2)^2 \qty(\frac{D}{2}-\frac{2}{3}) \qty(\frac{D}{2}-\frac{1}{3})
            \over
            \qty(\frac{D}{2}-\frac{3}{2})^2 \qty(\frac{D}{2}-1) \qty(\frac{D}{2}-\frac{1}{2})
    }
\end{equation}

The second simplification, is that as we will argue further, all
$\omega_{i}(D)$ for $i>1$ are zero. To see this, first let us
look at the asymptotic behavior of $F_i(D)$ at large $D$. Rewriting
eq.~\eqref{eq:fi} as an integral over invariants $s_{ij}$ gives
\begin{equation}
    \label{eq:fi-inv}
    F_i = S_\Gamma
    \left( \prod_{k=1}^{N-1} \frac{\Omega_{D-k}}{2} \right)
    \int
        \left (\prod_{l<m} \frac{\D s_{lm}}{2} \right)
        \big({\Delta_N}\big)^{\frac{D-N-1}{2}}
        \Theta\big(\Delta_N\big)\x
        \delta\left(1 - s_{1\dots N}\right)
        \frac{1}{D^{(i)}_1 ~ \dots ~ D^{(i)}_n},
\end{equation}
where $\Delta_N$ is the Gram determinant defined as
\begin{equation}
  \label{eq:gramdet}
  \Delta_N = \frac{(-1)^{N+1}}{2^N}
  \begin{vmatrix}
    s_{11} & s_{12} & \cdots & s_{1N} \\
    s_{12} & s_{22} & \cdots & s_{2N} \\
    \vdots  & \vdots  & \ddots & \vdots  \\
    s_{1N} & s_{2N} & \cdots & s_{NN}
  \end{vmatrix},
\end{equation}
and $\Omega_{k}$ is the surface area of a unit hypersphere in $k$-dimensional space
\begin{equation}
    \Omega_k = 2\pi^{\frac{k}{2}} \x \Gamma\left(\frac{k}{2}\right)^{-1}.
\end{equation}

If $\Delta_N(s_{ij})$ has a unique global maximum inside the integration region, we can
apply Laplace's method to eq.~\eqref{eq:fi-inv} and find its asymptotic as
\begin{equation}
    \label{eq:laplace-asy}
    F_i(D \to \infty) =
        S_\Gamma
        \left( \prod_{k=1}^{N-1} \Omega_{D-k} \right)
        \left( \Delta^{max}_N \right)^{\frac{D}{2}}
        \left( \frac{2\pi}{D} \right)^{\frac{1}{2}\left(\frac{N (N-1)}{2}-1\right)}
        \left( \mathcal{C}_i + \order{D^{-1}} \right),
\end{equation}
where $\mathcal{C}_i$ is a constant that depends on the location of the maximum
and the denominators $D^{(i)}_j$, but not on $D$.

The global maximum of $\Delta_N$ is reached when all $s_{ij}$ ($i\neq j$) are
identical and equal to $\frac{2}{N (N-1)}$. Geometrically this configuration
corresponds to the vectors $\vec{p}_i$ pointing to the vertices of a regular
$N$-hedron embedded into Euclidean space of $(N-1)$ dimensions. The maximum
value is then
\begin{equation}
    \label{eq:gramdet-max}
    \Delta^{max}_N = \frac{1}{N^{N} (N-1)^{N-1}},
\end{equation}
and explicitly we get
\begin{equation}
    \label{eq:fiasy}
    F_i(D \to \infty) =
        \pi^{\frac{7}{2}} 2^{\frac{25}{2}}
        \frac{\left(4^4 5^5\right)^{-\frac{D}{2}} \Gamma \left(\frac{3 D}{2}-3\right)}
        {D^{\frac{9}{2}}
            \Gamma \left(\frac{D-4}{2}\right)
            \Gamma \left(\frac{D-3}{2}\right)
            \Gamma \left(\frac{D-1}{2}\right)}
        \qty( \mathcal{C}_i + \order{D^{-1}} )
        \sim
        \frac{1}{D^{\frac{5}{2}}} \qty(\frac{3^3}{4^4 5^5})^\frac{D}{2}.
\end{equation}
It follows that all $F_i(D)$ have identical asymptotic behavior up
to a constant $\mathcal{C}_i$. As a confirmation, it can be shown that
eq.~\eqref{eq:fiasy} is asymptotically the same expression as we had for $F_1$ in
eq.~\eqref{eq:f1}.

Now we can compare this asymptotic for $F_i(D)$ to the asymptotic
for $H_i(D)$ from eq.~\eqref{eq:hi-rat-sol}. Indeed, for $H_{28}(D)$
from eq.~\eqref{eq:h28-rat-sol} we have:
\begin{equation}
    H_{28}(D) \sim \frac{1}{D^{\frac{1}{2}}} \qty(-\frac{3^3}{4^3})^{\frac{D}{2}}.
\end{equation}

This grows asymptotically exponentially faster than eq.~\eqref{eq:fiasy}.
In fact, all $H_i(D)$ for $i>1$ grow exponentially faster than $F_i(D)$,
and since $F$ and $H$ are connected via eq.~\eqref{eq:sol-hom-part},
this can only happen if the corresponding periodic functions
$\omega_i(D)$ are zero.


Thus, to find $F_i(D)$ we only need to find $R_i(D)$, the inhomogeneous solutions
to eq.~\eqref{eq:drr_i}. We compute them as a series in $\eps=(4-D)/2$ using
\dream with 2000-digit accuracy, and then restore the analytical form of the
series coefficients in terms of MZVs using the PSLQ method~\cite{FBA99}. This
way we obtain the analytical result for all master integrals up to MZVs of
weight 12 using the corresponding bases from~\cite{Fur03} and the \summertime
package~\cite{LM15} for their numerical evaluation. Corresponding expressions
are presented in the auxiliary files on the arXiv.

\section{Crosschecks}
\label{sec:4}

\subsection{Four-Particle Integrals}

As the first consistency check of our method we reproduce results
for four-particle phase-space integrals reported in~\cite{GGH03}.
We perform all the steps described in Section~\ref{sec:2}.
Generating the IBP rules with the help of \litered and then
proceeding with \dream we obtain the final result with 2000-digit
accuracy. The series reconstructed with PSLQ, and truncated to
MZV weight 6 are (using the original notation, and omitting
$S_\Gamma$ and $q^2$ factors):
\begin{align}
    R_6 \equiv \parbox{100pt}{\begin{axopicture}(100,40)
\Line(10,20)(20,20)
\Line(80,20)(90,20)
\den(20,20,50,40)
\cut(50,40,80,20)
\cut(20,20,50,0)
\den(50,0,80,20)

\arc(30,20,27,-45,45)
\arc(70,20,27,135,215)
\end{axopicture}
} &= -1+\z(2)
+\eps \Bigg(
    -12
    +5\z(2)
    +9\z(3)
    \Bigg)
+\eps^2 \Bigg(
    -91
    +27\z(2)
\\ \nn &
    +45\z(3)
    +\frac{61}{5}\z(2)^2
    \Bigg)
+\eps^3 \Bigg(
    -558
    +161\z(2)
    +197\z(3)
    +61\z(2)^2
\\ \nn &
    -80\z(3)\z(2)
    +207\z(5)
    \Bigg)
+\eps^4 \Bigg(
    -3025
    +939\z(2)
    +897\z(3)
\\ \nn &
    +\frac{1157}{5}\z(2)^2
    -400\z(3)\z(2)
    +1035\z(5)
    +\frac{288}{5}\z(2)^3
    -153\z(3)^2
\Bigg)
\\ \nn &
+ \order{\eps^5}
, \\
    R_{8,a} \equiv \parbox{100pt}{\begin{axopicture}(100,40)
\Line(10,20)(20,20)
\Line(80,20)(90,20)
\den(20,20,40,40)
\cut(40,40,60,40)
\den(60,40,80,20)
\den(20,20,40,0)
\cut(40,0,60,0)
\den(60,0,80,20)
\cut(40,40,60,0)
\dot(50,20)
\cut(60,40,40,0)
\end{axopicture}
} &= \frac{5}{\eps^4}
-\frac{40 \z(2)}{\eps^2}
-\frac{126 \z(3)}{\eps}
+14 \z(2)^2
+\eps \Bigg(1008 \z(2) \z(3)-1086 \z(5)\Bigg)
\\ \nn &
+\eps^2 \Bigg(-\frac{272}{7} \z(2)^3+1602 \z(3)^2\Bigg)
+ \order{\eps^3}
, \\
    R_{8,b} \equiv \parbox{100pt}{\begin{axopicture}(100,40)
\Line(10,20)(20,20)
\Line(80,20)(90,20)
\den(20,20,40,40)
\den(40,40,60,40)
\cut(60,40,80,20)
\cut(20,20,40,0)
\den(40,0,60,0)
\den(60,0,80,20)
\cut(40,40,60,0)
\dot(50,20)
\cut(60,40,40,0)
\end{axopicture}
} &= \frac{3}{4 \eps^4}
-\frac{17 \z(2)}{2 \eps^2}
-\frac{44 \z(3)}{\eps}
-\frac{183}{5} \z(2)^2
+\eps \Bigg(376 \z(2) \z(3)-790 \z(5)\Bigg)
\\ \nn &
+\eps^2 \Bigg(-\frac{19088}{105} \z(2)^3+698 \z(3)^2\Bigg)
+ \order{\eps^3}
.
\end{align}

These values are in a complete agreement with the known results,
and we have included the series up to MZV weight 12 among the
auxiliary files on the arXiv.

\subsection{Numerical Verification}

\begin{table}[h!]
\centering
\begin{tabular}{ rrrrrrr }
 $i$   & \multicolumn{3}{r}{Numerical results}& \multicolumn{3}{r}{Analytic results} \\
\hline
 &      $D=4$ &      $D=6$ &      $D=8$ &      $D=4$ &      $D=6$ &      $D=8$ \\
\hline \\ [-1em]
  2 &         -- &  1708(2)00 &   4699(1)0 &         -- &  171085.62 &  47000.531 \\
  3 &  3.7823(4) &  3.1704(2) &  3.0221(1) &  3.7823736 &  3.1704486 &  3.0221118 \\
  4 &         -- &  1504.7(8) &   725.3(1) &         -- &  1504.4507 &  725.26806 \\
  5 &         -- &  1007.4(5) &  580.80(9) &         -- &  1007.5235 &  580.76347 \\
  6 &         -- &  6191(5)00 &  14496(6)0 &         -- &  619633.25 &  144975.32 \\
  7 &   46.46(4) &  18.533(2) &  15.205(1) &  46.435253 &  18.532303 &  15.205538 \\
  8 &         -- &   2031(2)0 &    5357(2) &         -- &  20297.189 &  5355.3611 \\
  9 &         -- &    4313(3) &  2406.7(4) &         -- &  4312.8823 &  2406.7943 \\
 10 &  10.436(2) &  7.1508(5) &  6.5093(3) &  10.435253 &  7.1507477 &  6.5092878 \\
 11 &   228.8(1) &   62.67(1) &  47.663(4) &  229.11836 &  62.667046 &  47.663194 \\
 12 &         -- &  157.34(4) &  102.26(1) &         -- &  157.33521 &  102.26408 \\
 13 &         -- &   13729(8) &    4000(1) &         -- &  13732.166 &  4000.2779 \\
 14 &         -- &  268.46(8) &  172.80(2) &         -- &  268.45969 &  172.79805 \\
 15 &         -- &   6322(6)0 &   16048(5) &         -- &  63316.356 &  16049.857 \\
 16 &         -- &   4414(3)0 &   12952(4) &         -- &  44117.898 &  12951.443 \\
 17 &         -- &  1243.4(7) &   709.6(1) &         -- &  1243.1369 &  709.52840 \\
 18 &         -- &  3002(2)00 &   5899(3)0 &         -- &  300402.99 &  58965.517 \\
 19 &         -- &  4982(4)00 &  10637(4)0 &         -- &  498329.79 &  106357.81 \\
 20 &         -- & 2360(2)000 &  5777(2)00 &         -- &  2362594.9 &  577686.64 \\
 21 &         -- &   6312(5)0 &   20642(5) &         -- &  63147.876 &  20642.071 \\
 22 &         -- &  8402(7)00 &  24407(8)0 &         -- &  840453.94 &  244075.75 \\
 23 &         -- & 1443(1)000 &  4556(1)00 &         -- &  1443198.3 &  455543.43 \\
 24 &         -- &  1391(1)00 &   3997(1)0 &         -- &  139263.92 &  39966.878 \\
 25 &         -- &  3347(3)00 &   8526(3)0 &         -- &  335128.10 &  85254.217 \\
 26 &  25.563(6) &  15.376(1) & 13.6042(7) &  25.564747 &  15.376404 &  13.604247 \\
 27 &         -- &   697.3(3) &  397.84(6) &         -- &  697.18948 &  397.83514 \\
 28 &   143.9(1) &  52.855(7) &  42.917(3) &  143.97886 &  52.853837 &  42.917424 \\
 29 &         -- &   4409(3)0 &   13702(3) &         -- &  44117.898 &  13700.597 \\
 30 &         -- &   6327(6)0 &   16181(5) &         -- &  63316.356 &  16178.566 \\
 31 &         -- &   8955(8)0 &   19055(8) &         -- &  89611.062 &  19051.115 \\
\hline
\end{tabular}
\caption{%
Numerical results for the ratio $F_i/F_1$ with the corresponding uncertainties
(standard deviations) indicated in the parenthesis. Missing entries correspond
to divergent integrals.}
\label{table:numresults}
\end{table}

As another cross-check we have calculated the leading terms of the
$\epsilon$-expansion of $F_i$ numerically using the direct way: through
Monte-Carlo integration of eq.~\eqref{eq:fi} over the phase-space. While
such a technique cannot be easily applied to divergent integrals, we can
sidestep this issue by noting that our master integrals only suffer from
IR divergences that disappear already at $D=6$. This way we can check
several leading terms of the expansion at $D=4-2\epsilon$ by calculating the
corresponding integrals in $D\ge6$ since both are connected by dimensional
recurrence relations.

To calculate a finite integral of the form eq.~\eqref{eq:fi} we choose a
uniform mapping from a hypercube into momentum coordinates using an algorithm
similar to RAMBO~\cite{KSE85} but extended into arbitrary (integer) $D$. Then we
calculate the integrand from scalar products of the momenta, and finally we
integrate over the hypercube using the Vegas~\cite{Lepage78} implementation
from \textsc{Cuba}~\cite{Hah04}.

Note that although the integrals we are calculating are finite, the integrands
are not. Exposing an integration algorithm like Vegas to such infinities may
lead to unpredictable behaviour, so as a precaution we choose to regulate these
infinities by adding a small parameter $\alpha$ to the denominator of the
integrand, and then to calculate the integral with progressively smaller values
of $\alpha$ (from $2^{-30}$ to $2^{-100}$), checking if convergence was reached
afterwards.

The results of this method are summarized in Table~\ref{table:numresults}, and
show good agreement between numerical and analytic results. Our integration
program is written in C using the GNU Scientific Library~\cite{GSL09} and
\textsc{Cuba}. Its source code can be found at \url{https://hg.tx97.net/rambo},
and also in the auxiliary files on the arXiv. With a requested accuracy of $0.1\%$
the complete integration takes less than two days on a 12-core machine, with
each integration taking between a minute and two hours.

An alternative to the RAMBO-like algorithm, we have also implemented
the generation of phase space points by a sequence of two-particle
decays. This method is computationally faster, but does not give
a flat distribution of points over the phase space. Still, the
convergence of Vegas integration with this method appeared
to be as good as with the RAMBO version. We do not supply a
corresponding table of results, because it is very similar to
Table~\ref{table:numresults}.

We would like to note that such numerical integration is commonly done
differently, e.g. via the method sector decomposition~\cite{BH2000}.
To apply that method to phase space integrals one needs to
parameterize the phase space by a suitable rational mapping
from a hypercube, and for five particles we found this to be
challenging, whereas a four-particle phase-space mapping is known
from~\cite{GGH03}.

\section{Conclusions and outlook}
\label{sec:5}
In this article we present analytical expressions for five-particle
phase-space integrals expressed in terms of multiple
zeta values up to weight 12. The results are calculated using
dimensional recurrence relation method with a 2000-digit accuracy
using the \dream package, and analytically restored via the
PSLQ algorithm. We also present a computer code for the numerical
integration of phase-space integrals in a higher-number of
dimensions that has been used to cross-check the obtained
results. The approach presented here shows excellent performance
for calculating single-scale integrals without ultra-violet
divergences and can be easily applied to other problems of this
kind.

Further work in this direction includes the calculation of
multi-scale phase-space integrals needed for the extraction of
time-like splitting functions, with the presented integrals
used as boundary conditions of differential equations. Also of
interest is the calculation of the remaining unknown integrals
with 2-, 3- and 4-particle cuts appearing in the optical theorem
for the four-loop propagator.

\acknowledgments
We are thankful to Sven Moch for numerous discussions and helpful
suggestions concerning this work, and for proofreading this
article. We were pleased to use \texttt{Axodraw2}~\cite{CV16}
to draw diagrams for this article.

This work was supported in part by the German Research
Foundation DFG through the Collaborative Research Centre No.\ SFB~676
\textit{Particles, Strings and the Early Universe: the Structure of Matter and
  Space-Time}.

\bibliography{ps1to5}

\end{document}